\documentstyle[hip-artc]{article}  

\volnumber{5}  \edyear{1997}  \frompage{000} \topage{000}                
\recrevdate{1 January 1996; revised version 1 January 1997} 			 
\def\rightmark{How dense does parton matter get in Pb + Pb Collisions
at the CERN SPS ?}
\def\leftmark{K. Geiger and B. M\"uller}
 
\title{How dense does parton matter get in Pb + Pb Collisions
at the CERN SPS ?}
\authors{
{\twerm K. Geiger$^1$ and B. M\"uller$^{2}$ %
}\\[2.812mm]
{\normalsize
\hspace*{-8pt}$^1$ 
Brookhaven National Laboratory, Upton, NY 11973, USA\\
[0.2ex] 
\hspace*{-8pt}$^2$ 
Department of Physics, Duke University,
Durham, NC 27708-0305, USA
}}
 
\abstract{
We examine the  qualitative features of parton production
through materialization in heavy-ion collisions
within perturbative QCD, and estimate the magnitude of the
resulting parton density created during the early stage of the collisions.
The implications for ``anomalous'' $J/\psi$ suppression observed in
Pb+Pb collisions at the CERN SPS are discussed.
We argue that the $A$-dependence of absorption of $J/\psi$ by (partonic)
comovers is steeper than assumed in most phenomenological models,
because the absorption process is dominated by quasi-perturbative
QCD interactions. Our argument is supported by results recently
obtained in the framework of the parton cascade model. We predict
significant ``anomalous'' suppression for Pb+Pb collisions at the
CERN-SPS, but not for S+U collisions.
\\
\\
\\
\leftline{Pacs No.: 25.75.-q, 12.38.Mh}
} 

\input{epsf}
 
\begin{document}
 
\maketitle

\section{Introduction}

The creation of high-density QCD matter consisting of quarks
and gluons that are being liberated in ultra-relativistic heavy-ion collisions,
is one of the most fundamental issues for investigating
the new physics regime of deconfined parton matter and the quark-gluon plasma.
In this context,
the question whether the NA50 data \cite{NA50} on $J/\psi$ production
in Pb + Pb collisions at the CERN-SPS signal the presence of a novel
nuclear suppression mechanism is a matter of heated controversy.  Some
theoretical papers \cite{BO96,KLNS,Wong} have argued that the additional 
suppression observed in Pb + Pb, compared with S + U, is ``anomalous'', 
i.e. provides evidence of a new mechanism, most likely dissociation
of the $c\bar c$ quark pair in a dense deconfined medium, or
quark-gluon plasma \cite{MS85}.  Others \cite{ACXX,AC97,BC97,Gavin} 
have argued that ``conventional'' mechanisms, such as absorption of 
$J/\psi$ on light hadrons emitted at the same rapidity, can explain 
the increased suppression in Pb + Pb collisions.

It is not our goal here to add another phenomenological analysis of
the NA50 data to the existing ones.  Rather, we will make an attempt
to analyze on the basis of the established concepts of perturbative
QCD, whether Pb + Pb collisions ($P_{beam}=158$ $A$ GeV) are indeed so 
different from S + U collisions at almost the same energy 
($P_{beam}=200$ $A$ GeV) that
the former system might result in the formation of a thermalized, color-deconfined
state whereas the latter does not.  Our reasoning is based partly on
qualitative arguments first annunciated a decade ago \cite{BM87} and
partly on quantitative results \cite{GS97} recently obtained in the
framework of the parton cascade model of high-energy nuclear collisions
\cite{KM92,PRep95}.

\section{$A$-dependence of parton production}

We begin with the qualitative arguments.  Energy deposition at large
transverse momentum $p_\perp$ in hadronic collisions can be described in
the framework of perturbative QCD through jet or mini-jet production.
The crucial question is how low in $p_\perp$ this description remains
reliable.  There is an increasing amount of evidence \cite{Rev} that the
perturbative domain of QCD extends down to (spacelike) momenta of the
order of 1 GeV/$c$.  Contrary to hadron collisions at presently accessible
energies, in heavy-ion collisions 
the applicability of perturbative methods involves an additional scale
namely the parton density in the colliding nuclei.
For large nuclei even at moderately small values of the Bjorken
variable $x$, the
Lorentz contraction of the nucleons may lead to overlapping
partons with the net effect that  parton densities appear to be 
suppressed as compared to isolated hadrons.
Hence, not only the momentum transfer $p_\perp$ of mini-jet production,
but also the nuclear mass number $A$ determine the perturbative regime
of parton dynamics.

In order to estimate the $A$-dependence of the density of partons produced 
in heavy nucleus collisions and to pin down the point of
critical density where the parton densities begin to deviate,
we follow ref.~\cite{BM87} by considering the number of
mini-jets produced in a nucleus-nucleus collision $A+B$ at impact parameter
$b$ \cite{minijet}:
\begin{equation}
x {dN(b)\over dp_\perp^2dx} = 2 {\cal T}_{AB}(b) 
\;\sum_{ab}\,\int_{{4p_\perp^2\over xs}}^1 {dx'\over
x'} \,x'f_{a/A}(x',p_\perp^2) \;xf_{b/B}(x,p_\perp^2)\; {d\hat\sigma_{ab}\over dp_\perp^2}
\label{e1}
\end{equation}
where ${\cal T}_{AB}(b) = \int d^2s T_A(s) T_B(b-s)$ with the nuclear profile
functions $T_A$ and $T_B$ normalized to unity.  
The functions $f_{a/A}$ and $f_{b/B}$  are the 
distribution of partons $a,b = g, q_i, \bar{q}_i$ in the colliding nuclei
$A$ and $B$, which depend on the energy fractions $x, x'$ and the transverse
momentum $p_\perp$ of the partonic scattering with cross-section
$\hat\sigma_{ab}(x,x',p_\perp^2)$.
For purpose of simplicity, we consider now symmetric collision systems, i.e.
$A = B$, but the arguments below generalize straightforwardly to
asymmetric systems $A \ne B$.

To proceed, we recall  that for sufficiently large values of $p_\perp$ and 
not too small values 
of $x$, the nuclear parton distribution functions $f_{a/A}$ 
are approximately given by 
\begin{equation}
f_{a/A}(x,p_\perp^2) \;\approx\; A\; f_{a/N}(x,p_\perp^2), \label{e2}
\end{equation}
where $f_{a/N}$ is the number density of parton species $a$ in the 
nucleon $N$.  However, as $x$ and $p_\perp$ are decreased, the 
parton density in a heavy nucleus is
increasingly reduced (``shadowed'') because partons from different
nucleons overlap spatially and begin to fuse.  
Since the small-$x$ regime is strongly dominated by the rise of the
gluon density with quarks being negligible, 
the behavior of the gluon distribution $f_{g/N}(x,p_\perp^2)$ 
with decreasing $x$ and $p_\perp$ may be viewed as 
representative indicator for shadowing effects to become significant.
For gluons, the characteristic
momentum $p_{\rm crit}$ where this happens is determined by 
the number of gluons per unit area in the transverse plane of the
colliding nuclei at this momentum scale \cite{BM87},
\begin{equation}
p_{\rm crit}^2 = {9\over 16} \pi\alpha_s \,C_A\; {Ax f_{g/N}(x,p_{\rm crit}^2)
\over R_A^2}, \label{e3}
\end{equation}
where $C_A = 3$ is the value of the quadratic Casimir operator of
SU(3) in the adjoint representation, and $R_A = 1.2\,A^{1/3}$ fm is 
the nuclear radius.  Notice that $p_{\rm crit}^2$ enters as argument of 
$f_{g/N}$ on the right side; therefore (\ref{e3}) is a non-trivial implicit 
equation.  Nevertheless, we may get a rough estimate of the condition 
(\ref{e3}) by taking for instance $p_{\rm crit} = 1$ GeV and
$x f_{g/N}(x,1 \mbox{GeV}^2) = 3$ at $x\approx 0.1$ with 
$\alpha_s \approx 0.3$ at $p_{\rm crit}^2$, in which case one finds 
that shadowing corrections would become significant only for nuclei 
with $A > 240$.

Since $R_A\sim A^{1/3}$, the critical transverse momentum grows 
approximately as $p_{\rm crit} \sim A^{1/6}$.  The effect of shadowing 
is to reduce the number of mini-jets with transverse momenta $p_\perp < 
p_{\rm crit}$.  For larger values of $p_\perp$, the mini-jet 
yield grows as $A^{4/3}$, because $f_{g/A} \sim A \,f_{g/N}$ and 
${\cal T}_{AA}(b) \sim R_A^{-2} \sim A^{-2/3}$.  
However, approximating the glue-glue cross section by its small angle limit
\begin{equation}
{d\hat\sigma_{gg}\over dp_\perp^2} \approx {(\alpha_sC_A)^2\pi\over 2p_\perp^4},
\label{e4}
\end{equation}
one finds that the integrated mini-jet yield for all momenta $p_\perp >
p_{\rm crit}$ only grows linearly with $A$:
\begin{equation}
\int_{p_{\rm crit}^2}^{\infty} dp_\perp^2 \; x \; {dN(b) \over dp_\perp^2dx}
\sim A^{4/3} {1\over p_{\rm crit}^2} \sim A. \label{e5}
\end{equation}
This agrees with the result of Blaizot and Mueller \cite{BM87}, who
found that for two heavy nuclei the total yield of gluon mini-jets per
unit of rapidity is
\begin{equation}
{dN(b) \over dy} \;= \;2Ax f_{g/N}\; \frac{{\cal T}_{AA}(b)}{{\cal T}_{AA}(0)}
. \label{e6}
\end{equation}
The same considerations apply when the mini-jet production is mostly
due to quark-quark or quark-gluon scattering, because the differential
parton cross sections all have the same functional form (\ref{e4})
at small angles.

To summarize, we have the following picture describing the mini-jet
production at central rapidity in nucleus-nucleus collisions:  Above a
critical transverse momentum $p_{\rm crit}$, which itself grows as
$A^{1/6}$, the mini-jet multiplicity grows like $A^{4/3}$.  The total
mini-jet yield, as well as the multiplicity in the saturation regime
below $p_{\rm crit}$, only grows linearly with $A$.  This result
implies that the area density of mini-jets grows as $A^{2/3}$ above
$p_{\rm crit}$, but only as $A^{1/3}$ below $p_{\rm crit}$.

\section{Implications for $J/\psi$ suppression}

We now return to the problem of the observed strong suppression of
$J/\psi$ production in Pb + Pb collisions.  In particular, we want to
address the question of the $A$-dependence of the absorption of the
$J/\psi$ by comoving produced hadronic matter.  In the framework
developed by Bhanot and Peskin \cite{BP80}, interactions between light
hadrons and deeply bound, heavy quarkonium status, such as the
$J/\psi$, are mediated by short-range color dipole interactions.  The
relevant momentum transfers $p_\perp^2$ for these interactions lie in the
perturbative domain of QCD, $p_\perp^2 \ge 1\; {\rm GeV}^2$.

The key element in the following arguments is the fact that,
for most nuclear collision systems
at the relatively low center-of-mass energy of the CERN-SPS
experiments, perturbative mini-jet production is largely
due to quark-quark scattering, because the typical value of
Bjorken-$x$ probed is $\langle x \rangle 
\, \lower3pt\hbox{$\buildrel >\over\sim$}\,0.1$, where the gluon density 
is small.  This implies that shadowing effects are negligible,
i.e. the relation (\ref{e2}) holds.  Furthermore, 
according to (\ref{e3}) the critical momentum $p_{\rm crit}$
lies below the perturbatively accessible range of momenta
$p_\perp \ge 1$ GeV at the SPS energy, and may just begin to reach into
it for the Pb + Pb system.  
If this is correct, then {\it all} mini-jet production involving momenta
$p_\perp \ge 1$ GeV lies safely above $p_{\rm crit}$, where eq. (\ref{e2})
holds.
As immediate consequence, the density of
comovers which can effectively interact and absorb the $J/\psi$ at the
SPS grows like $A^{2/3}$, or $(A_1A_2)^{1/3}$ in asymmetric
collisions.  This is a {\it much stronger $A$-dependence} than naively
expected and embodied in most comover suppression models \cite{Gavin}.  
It also implies a {\it stronger impact parameter or $E_T$-dependence} of 
comover suppression than predicted by existing models.

Quantitative support for our speculations comes from recent calculations 
\cite{GS97} of secondary particle production in the parton cascade
model \cite{Klaus}.  This model is based on perturbative parton
scattering with dynamic, medium dependent cut-off mechanisms of the
infrared divergences of QCD.  The model also incorporates a density
dependent clustering mechanism for the parton-to-hadron transition.
The calculations predict that the energy density $\epsilon$ produced
by scattering partons (mini-jets) at central rapidity grows by a factor
of more than 2 between S + U and Pb + Pb, compatible with the scaling
law $\epsilon\sim(A_1A_2)^{1/3}$, but much faster than expected in
the usual Glauber model approach which predicts $\epsilon\sim
(A_1^{1/3}+A_2^{1/3})$, in which case $\epsilon$ increases only by 30 $\%$.

In Pb + Pb collisions the initial partonic energy density $\epsilon_p$
is predicted to reach 5 GeV/fm$^3$ at times $\tau < 1$ fm/$c$ in the
comoving reference frame (see Fig. 1).  For a thermalized gas of free 
gluons and three flavors of light quarks this corresponds to an initial
temperature $T_i \approx 230$ MeV, clearly above the critical
temperature $T_c\approx 150$ MeV predicted by lattice-QCD calculations 
\cite{Blum} shown in Fig. 2.  On the other hand, the parton density
$\epsilon_p \approx$ 2 GeV/fm$^3$ predicted for S + U collisions (Fig. 1), 
just lies at the upper end of the transition region in the equation of
state from lattice-QCD.

The temperature $T_D$ required for dissociation of the $J/\psi$ bound 
state due to color screening is known \cite{KS91} to be higher than
$T_c$, namely $T_D\approx 1.2 \,T_c = 180-200$ MeV.  The parton cascade model 
results are therefore compatible with the experimental finding that there
appears to exist no significant comover-induced suppression of $J/\psi$ in 
S + U collisions, but a large and strongly impact parameter dependent effect
is observed in Pb + Pb collisions.

In conclusion, we have argued that the density of comovers relevant
for $J/\psi$ suppression in nucleus-nucleus collisions grows more
rapidly as function of nuclear mass than assumed in previous studies
based on the Glauber model.  Recent results obtained in the framework
of the parton cascade model are therefore consistent with the interpretation
that the ``anomalous'' $J/\psi$ suppression observed in Pb + Pb
collisions is caused by color screening in a dense partonic medium.
The same calculations predict that there is no such effect in S + U
collisions.
It must be stressed that these model results do not contradict 
the experimental data collected at the SPS, as one might suspect,
since it is usually claimed that those are fully consistent with
the Glauber picture. 
In fact, the detailed study of particle spectra in Ref. \cite{GS97},
comparing model and experiment, shows that the data are compatible
with a parton cascade picture. In particular, the magnitude
of transverse energy production visible in the final-state hadron spectra,
is in fair agreement, although the maximum achieved energy density
during the very early stage in Fig. 1 is substantially larger than
previously estimated.

\section*{Acknowledgement}
This work was supported in part by grants
DE-FG02-96ER40495 and DE-AC02-76H00016 from the U.S.~Department of Energy.

\vfill\eject

\newpage
\begin{figure}
\epsfxsize=420pt
\centerline{\epsfbox{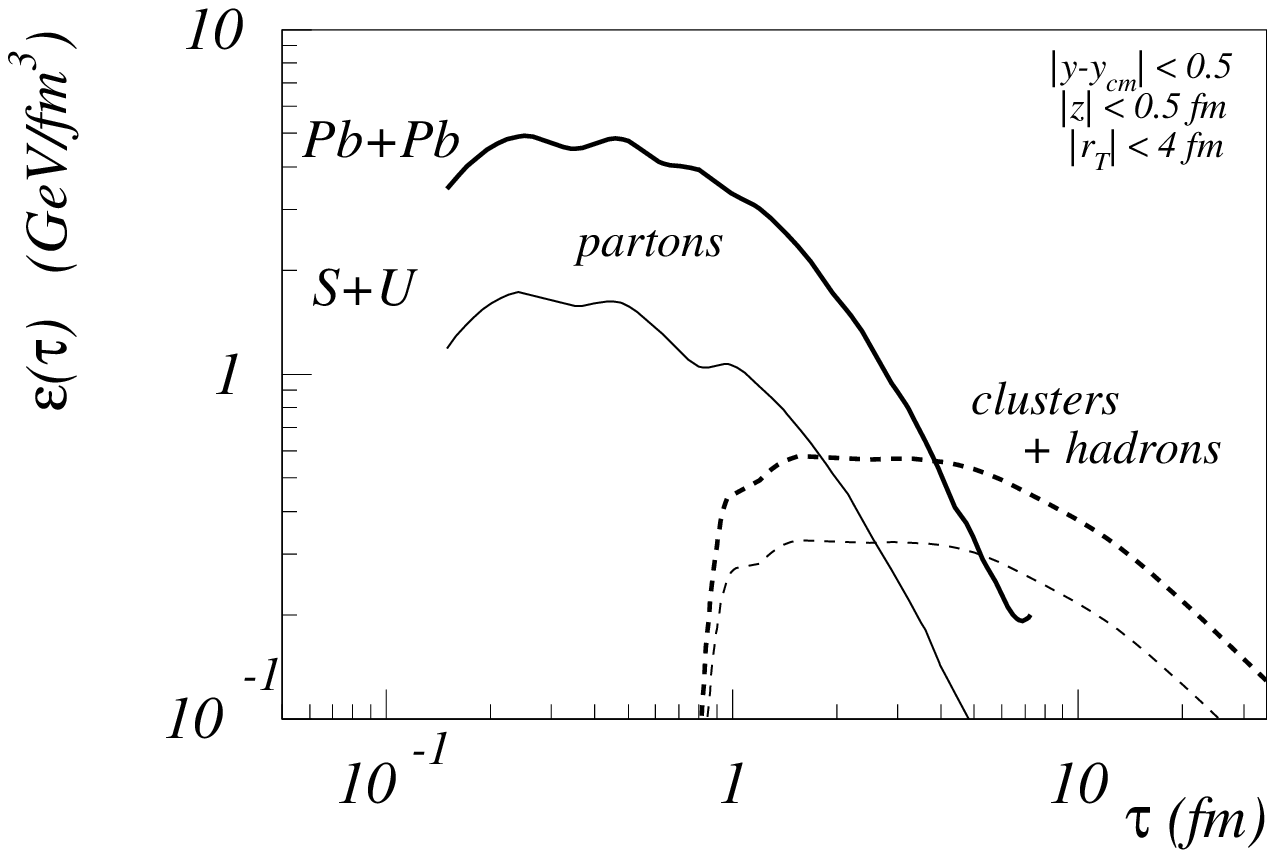}}
\caption{
Time evolution
of the energy density $\varepsilon$ of the partonic matter in the 
central slice of the collision systems S + U and Pb + Pb at CERN-SPS 
beam energies of 200 $A\cdot$GeV and 158 $A\cdot$GeV, respectively.
The model calculations were obtained with the MC implementation 
\protect\cite{Klaus} of the parton cascade model \protect\cite{KM92,PRep95}.
}
\label{fig:fig1}
\end{figure}
\newpage
\begin{figure}
\epsfxsize=420pt
\centerline{ \epsfbox{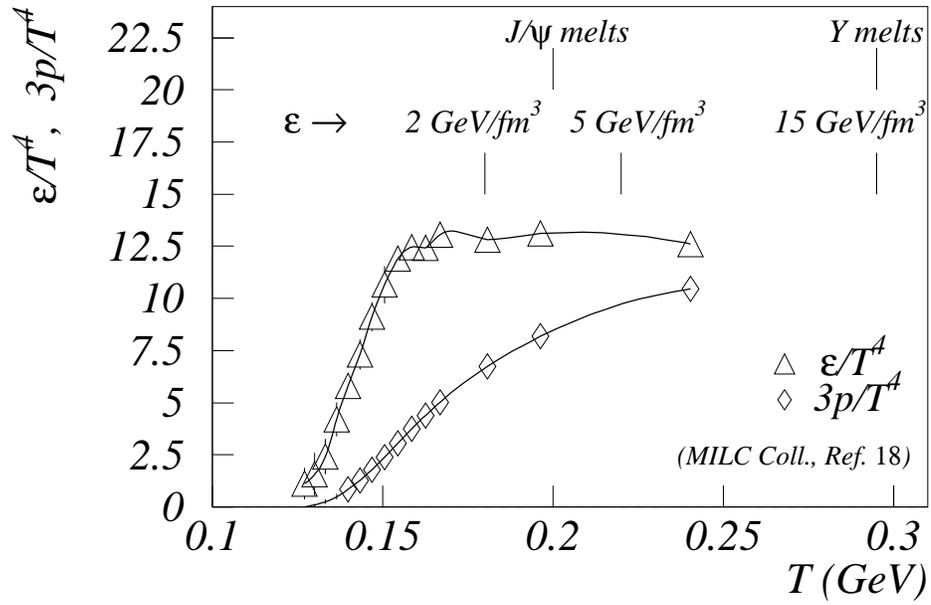}}
\caption{
Equation of state for two-flavor QCD from lattice calculations by the 
MILC collaboration \protect\cite{Blum}, showing energy density 
$\epsilon$ and pressure $p$ as a function of temperature
$T$. Indicated in the figure are also the energy densities 
corresponding to several values of the temperature $T$, as well as 
the location of the ''melting'' points of the $J/\psi$ and $\Upsilon$ 
states, where these bound states disappear due to dissociation 
\protect\cite{KS91}.
}
\label{fig:fig2}
\end{figure}
\end{document}